\documentclass[letter,traditabstract,longauth]{aa} 
\usepackage{graphicx}
\usepackage{natbib}
\usepackage{longtable}
\usepackage{rotating}
\usepackage[varg]{txfonts}
\usepackage{pdflscape}
\usepackage{multirow}
\usepackage{ulem}

\def\degr{\hbox{$^\circ$}}

\def\arcsec{\hbox{$^{\prime\prime}$}}

\newcommand{\microns}{$\mu$m }

\usepackage{txfonts}
\begin{document}
   \title{Mapping the column density and dust temperature structure of IRDCs with Herschel\thanks{Herschel is an ESA space observatory with science instruments provided by European-led Principal Investigators consortia and with important participation from NASA}}

 %  \subtitle{First glimpse into the column density and temperature structure of IRDCs}

   \author{N. Peretto
          \inst{1,2}, G.A. Fuller\inst{1}, R. Plume\inst{3}, L. D. Anderson\inst{4}, J. Bally\inst{5}, C. Battersby\inst{5}, M. T. Beltran\inst{6}, J.-P. Bernard\inst{8}, L. Calzoletti\inst{12}, A.M. DiGiorgio\inst{10}, F. Faustini\inst{12}, J. M. Kirk\inst{7}, C. Lenfestey\inst{1}, D. Marshall\inst{8}, P. Martin\inst{9}, S. Molinari\inst{10}, L. Montier\inst{8}, F. Motte\inst{2}, I. Ristorcelli\inst{8}, J.A. Rod\'on\inst{4},  H. A. Smith\inst{11}, A. Traficante\inst{12},  M., Veneziani\inst{12}, D. Ward-Thompson\inst{7}, L. Wilcock\inst{7}
          }

   \offprints{}

   \institute{\inst{1} Jodrell Bank Centre for Astrophysics,  School of Physics and Astronomy, The University of Manchester, Manchester, M13 9PL, UK\\
     \inst{2} Laboratoire AIM, CEA/DSM-CNRS-Universit\'e Paris Diderot, IRFU/Service d'Astrophysique, C.E. Saclay, Orme des merisiers, 91191 Gif-sur-Yvette, France\\
     \inst{3}Department of Physics \& Astronomy, University of Calgary, Canada\\
     \inst{4} Laboratoire d'Astrophysique de Marseille, UMR 6110 CNRS \& Universit\'e de Provence, France\\
     \inst{5} Center for Astrophysics and Space Astronomy, University of Colorado, Boulder, USA\\
     \inst{6} INAF, Osservatorio Astrofisico di Arcetri, Largo E. Fermi 5, 50125 Firenze, Italy\\
     \inst{7} School of Physics \& Astronomy, Cardiff University, Queens Buildings, The parade, Cardiff, CF24 3AA, UK\\
     \inst{8} Centre d'Etudes Spatiales des Rayonnements, 9 Avenue Colonel Roche, 31022 Toulouse, France\\
     \inst{9} INAF-Instituto Fisica Spazio Interplanetario, Via Fosso del Cavaliere 100, I-00133 Roma, Italy\\
     \inst{10} Department of Astronomy \& Astrophysics, University of Toronto, Toronto, Canada\\
     \inst{11} Harvard-Smithosonian Center for Astrophysics, 60 Garden Street, Cambridge, MA 02138, USA\\
     \inst{12} Dipartimento di Fisica, universit\`a di Roma 2, Tor vergata, Rome, Italy\\
     \inst{13} ASI Science Data Center, I-000444, Frascati (Rome), Italy\\
              \email{Nicolas.Peretto@manchester.ac.uk }}
              
%               \institute{Service d'Astrophysique, CEA-Saclay, France}

   \date{Received; accepted}
% \abstract{}{}{}{}{} 
% 5 {} token are mandatory
   \abstract{ Infrared dark clouds (IRDCs) are cold and dense reservoirs of gas potentially
     available to form stars. 
 Many of these clouds are likely to be pristine structures
representing the initial conditions for star formation.
  % aims heading (mandatory)
  The study presented here aims to construct and analyze accurate column
  density and dust temperature maps of IRDCs by using the first Herschel data from the Hi-GAL galactic plane survey. 
  These fundamental quantities, are essential for understanding  processes
  such as fragmentation in the early stages of the formation of stars in
  molecular clouds.
  % methods heading (mandatory)
%To map the temperature and column density we use the first Herschel
%  data obtained within the open time key program Hi-GAL which will map
%  the inner part of the galactic plane. Two tiles observed at 70, 160 with PACS and 
 %   250, 350, 500\microns\  with SPIRE, centered on $l=30\degr$ and $l=59\degr$
% encompass a total of more that 450
%  Spitzer IRDCs. 
 We have developed a
  simple pixel-by-pixel SED fitting method, which  accounts for the  background
emission. By fitting a grey-body function at each
  position, we recover the spatial variations  in both
    the dust column density and temperature  within the IRDCs. This
    method is applied to a sample of 22 IRDCs exhibiting a range of angular sizes and peak column densities. 
  % results heading (mandatory)
 Our analysis shows that the dust
  temperature decreases significantly within IRDCs, from background temperatures of 20-30~K  to
 minimum temperatures of 8-15~K within the clouds, showing that dense molecular clouds are not isothermal. Temperature gradients have most likely an important impact on the fragmentation of IRDCs. Local temperature 
 minima are strongly correlated  with column density peaks, which in a few cases reach N$_{H_2} = 1\times 10^{23}$~cm$^{-2}$,  identifying 
 these clouds as candidate massive prestellar cores. 
%GAF: I didn't like this conclusions section, so I rewrote it. 
%{The method \sout{we} developed {\bf here} \sout{allows us to have} access to
%  the true column density structure of IRDCs via the construction of
%  temperature maps. We therefore plan to generalize this method to a large
%  sample of IRDCs in order to get statistics on cloud fragmentation and their
%  thermal properties.}
%Applied to the entire Hi-GAL surveyWe showed that the technique developed here to the Hi-GAL multi-wavelength data provides a
 % method to map  temperature structure of IRDCs, necessary information to recover the true column density structure within IRDCs.
   Applying this technique to the full Hi-GAL data set  will provide important constraints on the fragmentation and thermal
  properties of IRDCs, and help identify hundreds of massive prestellar core candidates. 
}
  
    % context heading (optional)
  %{ppp} leave it empty if necessary 
%   {Infrared dark clouds are cold and dense reservoirs of gas potentially
%     available to form stars. The initial conditions of star formation are
%     likely to be still present in such pristine structures. }

\authorrunning{Peretto et al.}
\titlerunning{Mapping IRDCs with Herschel}
  
   \maketitle
%
%________________________________________________________________

\section{Introduction}

% Observations of star forming regions have been conducted for centuries now,
% always trying to see deeper into these accumulation of molecular gas where
% stars like to form. The main goal of star formation studies is to better
% understand what are the initial conditions of such a process, what determine
% the final mass of a star. The very first step toward understanding this is to
% know what are the physical properties of the gas out of which stars form, such
% as the column density and the temperature. While fundamental, these quantities
% are quite difficult to estimate, the main reason being that star forming
% regions are cold, 5-30~K, and therefore radiate energy away through dust in
% the far infrared/submillimeter wavelengths, frequency range barely accessible
% from the ground due to the opacity of the atmosphere.

One of the major issues in star formation is understanding the physical
  conditions, including the temperature and density structure, of the material
  out of which stars will form.  While fundamental, these quantities are
  difficult to determine. These regions are cold and therefore
  radiate energy away through dust at the far infrared/submillimeter
  wavelengths, a frequency range barely accessible from the ground because of the
  opacity of the atmosphere.  
 
For the first time, a high angular resolution view of the Universe from
  70 to 500 \microns\ is now becoming available thanks to the ESA Herschel Space
  Observatory \citep{pilbratt2010}. The frequency range provided by its two photometry instruments PACS \citep{poglitsch2010} and SPIRE \citep{griffin2010} perfectly matches the
emission peaks of young stellar objects and prestellar cores, the direct
progenitors of stars. It is  now possible to
  construct well constrained SEDs for these objects, and
therefore determine the spatial distributions of the
temperature and column density  
  towards a wide range of different objects.

Hi-GAL is the open time key project  to observe the inner part
($|l|\leq60\degr; |b| \leq1\degr$) of the Galactic plane with Herschel
\citep{molinari2010a}.  The sensitivity of this survey will allow the detection
of 10~M$_{\odot}$ molecular clouds at a distance of 4~kpc. A large number of
 diverse studies will  result from this survey
 covering scientific themes such as evolved
stars, diffuse interstellar medium, and star formation. In this  Letter,
we present the first  Herschel study analyzing the temperature and column
density structure of infrared dark clouds (IRDCs). These objects are
reservoirs of cold, dense molecular gas seen in extinction in the mid-infrared
against the strong background emission of the Galactic plane \citep[Fig.~\ref{sdc2col}]{perault1996,rathborne2006}.
This gas is only slightly processed by current star formation activity and most likely still
contains the initial conditions of star formation imprinted within them. Their
study can therefore illuminate the earliest stages of star formation. In Sect. 2 of this paper, we
describe the data, Sect. 3 focuses on the background properties. Section
4 presents the first results, while Sects. 5 and 6 summarize
the discussion and conclusions of the paper, respectively.

\section{Data}

This  analysis uses data  from  the
Hi-GAL Science Demonstration Phase (SDP)  which
consists of two tiles of $\sim 2.2\degr\times2.2\degr$  centered 
on $l=30\degr$ and $l=59\degr$. For each tile, the parallel mode of observations provides PACS  70/160~\microns and SPIRE
250/350/500~\microns  images. The SDP data were reduced using the
ROMAGAL data reduction software (see \citet{molinari2010b} for a complete description). 
The recommended calibration factors and offsets  were applied to the data  to set the zero flux level  \citet{bernard2010}.

\begin{figure}[!t!]
\hspace{-0.5cm}
\vspace{-0.0cm}
\includegraphics[width=9.5cm,angle=0]{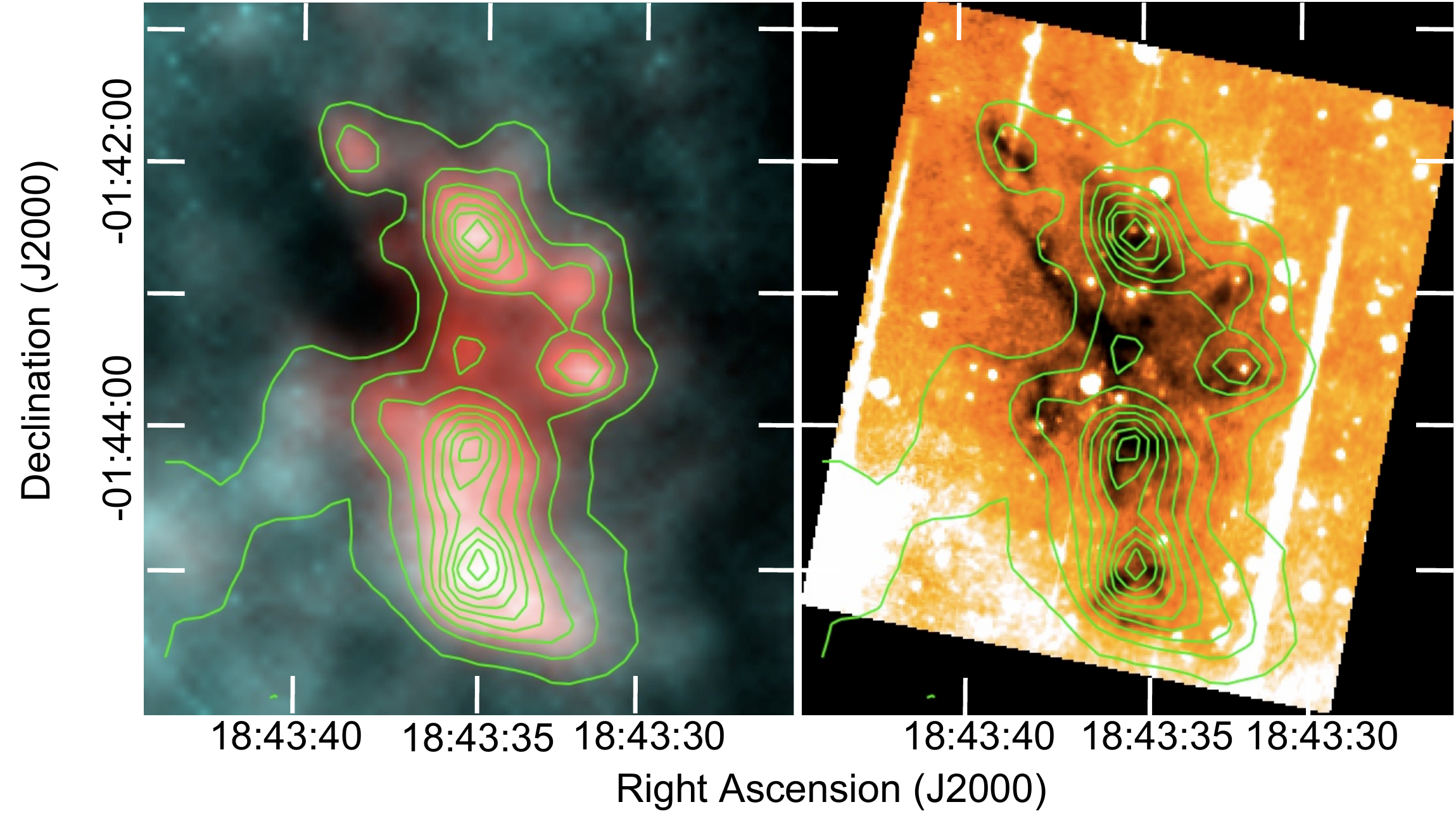}
%ss.eps}
%\vspace{-2cm}
\caption{ {\it (Left):} Two color image of one of the 22 IRDCs analyzed here. The blue is the 160~\microns emission while the red and the contours show the 250~\microns.{\it (Right):} Spitzer 8~\microns image of the same IRDC on which we have overlaid the Herschel 250~\microns contours.
  \label{sdc2col}}
\end{figure}

In total, \citet{peretto2009} catalogued $\sim450$ Spitzer IRDCs in
the 2 tiles of the Hi-GAL SDP fields, 75\% of these being in the
$l=30\degr$ tile. Checking each of these clouds by eye, we identified
80 to 90\% of these IRDCs seen in emission with Herschel at long
wavelengths, confirming that these are real clouds, not caused by artifacts
in the Spitzer 8~\microns background. For this study, we selected a
subsample of 22 IRDCs (see Fig.~\ref{sdc2col}),  with sizes
  ranging from 40\arcsec to 180\arcsec,\footnote{These sizes are those
    derived from the extinction and given in \citet{peretto2009}. The
    sources are systematically larger in emission because the Hi-GAL data
    probes lower column density material.} all large enough to contain
at least one 500~\microns beam (i.e., 36\arcsec) and with a column
density peak derived from extinction such as N$_{\rm H_2}^{\rm peak}
\geq 3\times10^{22}$~cm$^{-2}$, which is high enough to clearly stand
out of the Hi-GAL data background.  All images were resampled
  to a common pixel size of 2\arcsec.  We excluded 
% GAF 
%on purpose 
the W43 region because of its complex background structure
\citep{bally2010}. Our selected sample is clearly biased towards
rather large and massive IRDCs.

\section{Background emission towards the IRDC}

The equation  governing the emission $I_{\nu}$ towards an
IRDC in the optically thin case is:

\begin{eqnarray}
  I_{\nu} &=& B_{\nu}(T_d^{\rm IRDC})\left(1-e^{-\tau_\nu^{\rm IRDC}}\right) +
  I_{\nu}^{\rm back}e^{-\tau_\nu^{\rm IRDC}} + I_{\nu}^{\rm fore}\\
  &=& B_{\nu}(T_d^{\rm IRDC})\times\tau_{\nu}^{\rm IRDC} + I_{\nu}^{\rm bg},
\label{eqn}
\end{eqnarray}
where $B_{\nu}(T_{ d}^{\rm IRDC}$) is the Planck function at the temperature of
the IRDC averaged along the line of sight, $\tau_{\nu}^{\rm IRDC}$ is the
opacity of the IRDC at the corresponding frequency, and $I_{\nu}^{\rm bg}$ is the
combined background and foreground emission, which is approximately independent of $\tau_{\nu}^{\rm IRDC}$ in the optically thin
limit.  Equation~\ref{eqn}
shows that to calculate the opacity and temperature for images at a
range of wavelengths, the background emission $I_{\nu}^{\rm bg}$ at each
wavelength needs first to be determined.  

% GAF: Rewritten below
%When looking at large scale images obtained with Herschel (MADRID images) we
%quickly realize that large scale emission is everywhere and represents a
%significant fraction of the total flux in these images. Moreover, at long
%wavelengths, this flux adds up to the flux of the sources we are looking at
%since emission is, in a large majority, optically thin. Therefore,
%fluctuations in the background could be interpreted as fluctuations of the
%IRDC emission itself.  It is crucial to take this effect into account as best
%as we can.

\begin{figure}[!t!]
\hspace{-0.cm}
\vspace{-0.0cm}
\includegraphics[width=9.cm,angle=0]{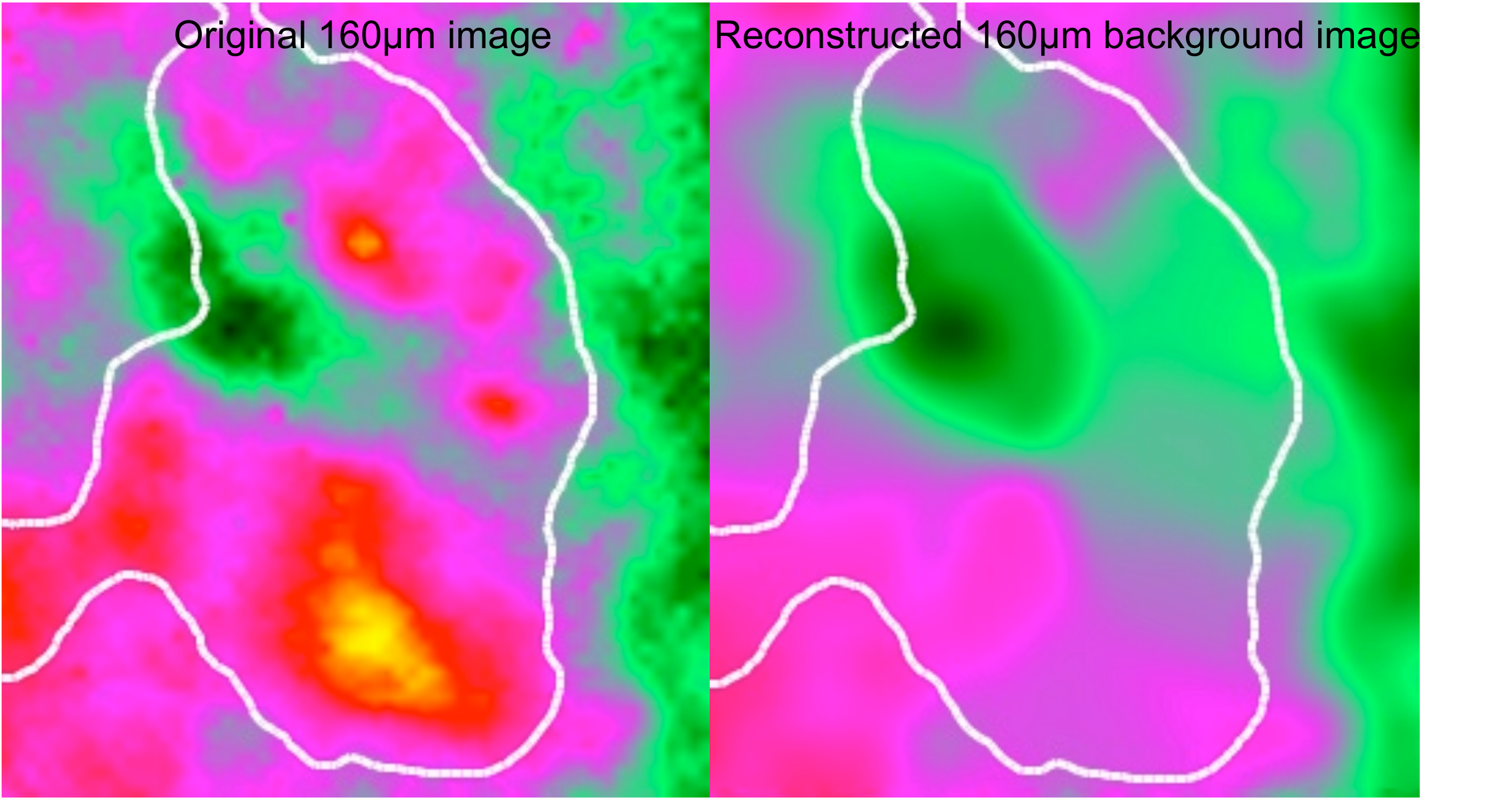}
%ss.eps}
%\vspace{-2cm}
\caption{Comparison of the original 160~\microns image (left) with the
  corresponding reconstructed background image (right). The intensity scale is
  the same for both images. The reconstructed background image has been smoothed to 36\arcsec, the common resolution used in for the SED fitting (cf Sects. 4 and 5). The contour corresponds to the limit between IRDC and background derived from the 500~\microns data.
\label{bg}}
\end{figure}

The full images show that a significant fraction of the total flux is
from large-scale extended emission \citep{molinari2010b}. It is
important to account for this emission to reliably map the IRDC
emission because fluctuations in this background could otherwise be
interpreted as structure in the IRDC. However defining the background
structure is not trivial.  Since the cirrus noise decreases towards
longer wavelengths \citep{gautier1992}, we used the 500~\microns
image to define the boundaries of each IRDC  (The 350\microns
  data, which are very similar but have higherr angular resolution,
  were also used to confirm the boundary of each IRDC.)  The pixel
flux distribution at 500~\microns toward IRDCs shows in most cases a
well defined peak at low fluxes and extended tail at larger fluxes
corresponding to the IRDC. The level to separate the background and
IRDC was chosen to be one standard deviation above the peak value, the
standard deviation being calculated from only those pixels below the
peak. This procedure allows us to identify the pixels in the image
associated with the IRDC, and those pixels containing only background
emission (Fig.~\ref{bg}).  This procedure works well, even for
  small IRDCs, provided there are sufficient pixels to characterize
  the background emission.

   To take the background fluctuations into account, we reconstructed  at all wavelengths background images
   by  interpolating the background pixels at the position of the IRDC pixels. After experimenting with various
   interpolation methods, we decided to use an interpolation based on the nearest neighbors. However, the interpolation method has little impact on the
   global physical properties since flux uncertainties are dominated by cirrus noise at short wavelengths (cf. next paragraph) and calibration uncertainties at longer wavelengths.
   In Fig.~\ref{bg}, we show the reconstructed
  background images of the IRDC at 160~\microns as well as the original image
  on the same intensity scale. The most significant feature of this background
  image is the important decrease in intensity on the left part of the image
  (as also seen at 250~\microns; Fig.~\ref{sdc2col}), which is
  recovered well on the reconstructed background image.
   
\begin{figure}[!t!]
\hspace{-0.cm}
\vspace{-0.0cm}
\includegraphics[width=8.5cm,angle=0]{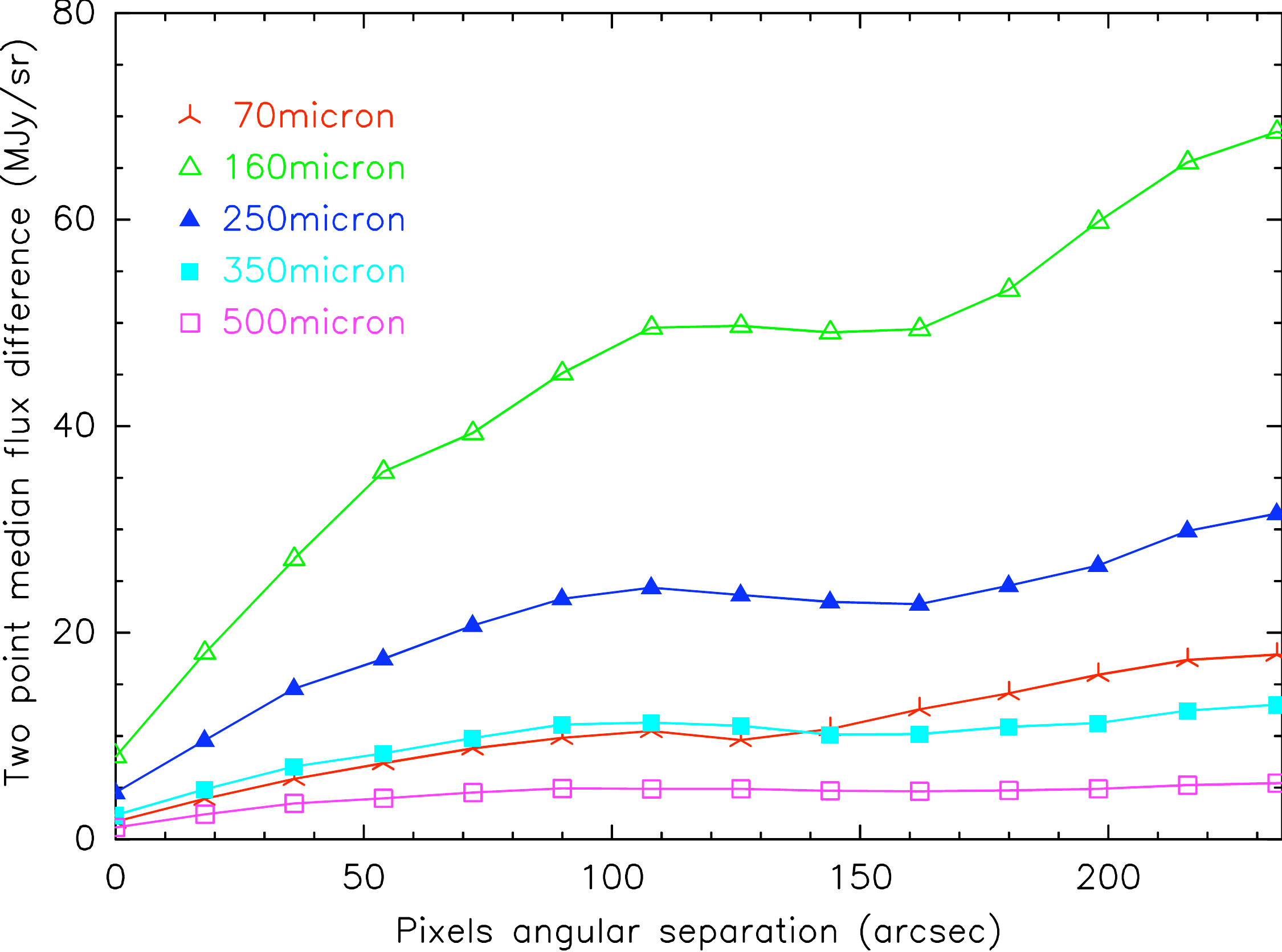}
%ss.eps}
%\vspace{-2cm}
\caption{Evolution of the flux difference between 2 background-only pixels as a function of their angular separation for the IRDC shown in Fig.~\ref{sdc2col} (after smoothing to a common 36\arcsec~resolution). This clearly shows that the background
is far more structured at 160~\microns than at 500~\microns, justifying our choice of using the latter to define the boundaries of the  IRDCs.
\label{uncert}}
\end{figure}

 Uncertainties in these background images are obviously higher  at the center of the IRDCs, where  pixels are further away from 
 a background pixel. To estimate this uncertainty, we computed a two point flux difference for all background pixels
 as a function of their angular separation, and this at each wavelength (Fig.~\ref{uncert}). As observed for all IRDCs, the uncertainties are much higher at 160~\microns than at 500~\microns, justifying our choice of choosing the latter for defining the IRDC boundaries.  When the uncertainties computed this way 
 are higher than the 20\% calibration uncertainty, we used them on the flux measurement when performing pixel by pixel SED fitting (cf. Sect. 4).

  The SED of  the background images
  were fitted to determine the properties of the dust responsible for the
  background emission (see Fig.~\ref{sed}).  With 5 data points per pixel (from 70 to 500
~\microns), we can constrain, in addition to opacity and dust
temperature, the spectral index $\beta$ of the specific dust opacity law,
 where $\tau_{\nu} \propto \nu^{\beta}$. Doing so, we find that the background dust temperature varies from 
 $\sim20$ up to $\sim30$~K depending on the position of the IRDC in the Galactic plane, in agreement
 with \citet{bernard2010}, while $\beta$ varies from 1.6 up to 2.3, consistent with \citet{boulanger1996}. 

%  {\bf GAF: This discussion would make more sense when talking about the mean
 %   dust properties in the background. Another question is how do the average
 %   properties of the dust derived from the background pixels and the
  %  reconstructed background towards the IRDC compare? I would hope that they
  %  are very similar.}

\section{Column density and dust temperature maps of IRDCs}

The reconstructed background images provide the $I_{\nu}^{\rm bg}$ term 
  in  Eq.~\ref{eqn} for each pixel of the IRDC.  The values of
  $\tau_{\nu}^{\rm IRDC}$ and $T_d^{\rm IRDC}$
 at each pixel are then determined by minimizing the $\chi^2$ between the
observed fluxes and those given by Eq.~\ref{eqn}. For this purpose, we used
the MPFITS package \citep{markwardt2009},
%\footnote{
as for the background SED fitting.

\begin{figure}[!t!]
\hspace{-0.cm}
\vspace{-0.0cm}
\includegraphics[width=9.cm,angle=0]{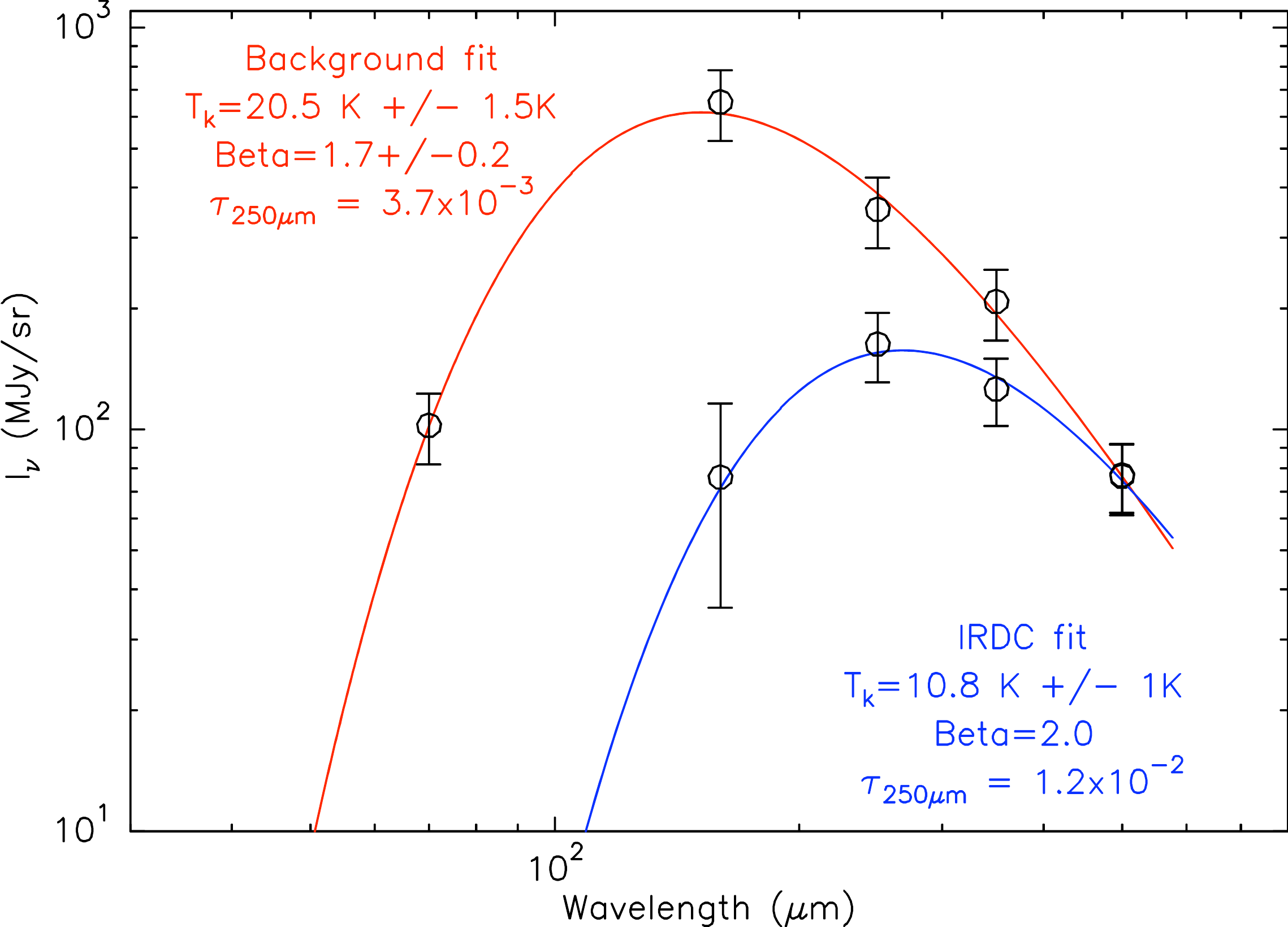}
%ss.eps}
%\vspace{-2cm}
\caption{SED fit of a background-only pixel of the IRDC shown in Fig.~\ref{sdc2col} (red) and the pixel at the column density peak (blue).
 The solid lines are the best fit to the data points. The uncertainties in the background are the 20\% calibration ones, while for the IRDC we took the larger of the calibration uncertainty and the ones described by Fig.~\ref{uncert}. 
 \label{sed}}
\end{figure}

\begin{figure}[!t!]
\hspace{-0.5cm}
\vspace{-0.0cm}
\includegraphics[width=9.cm,angle=0]{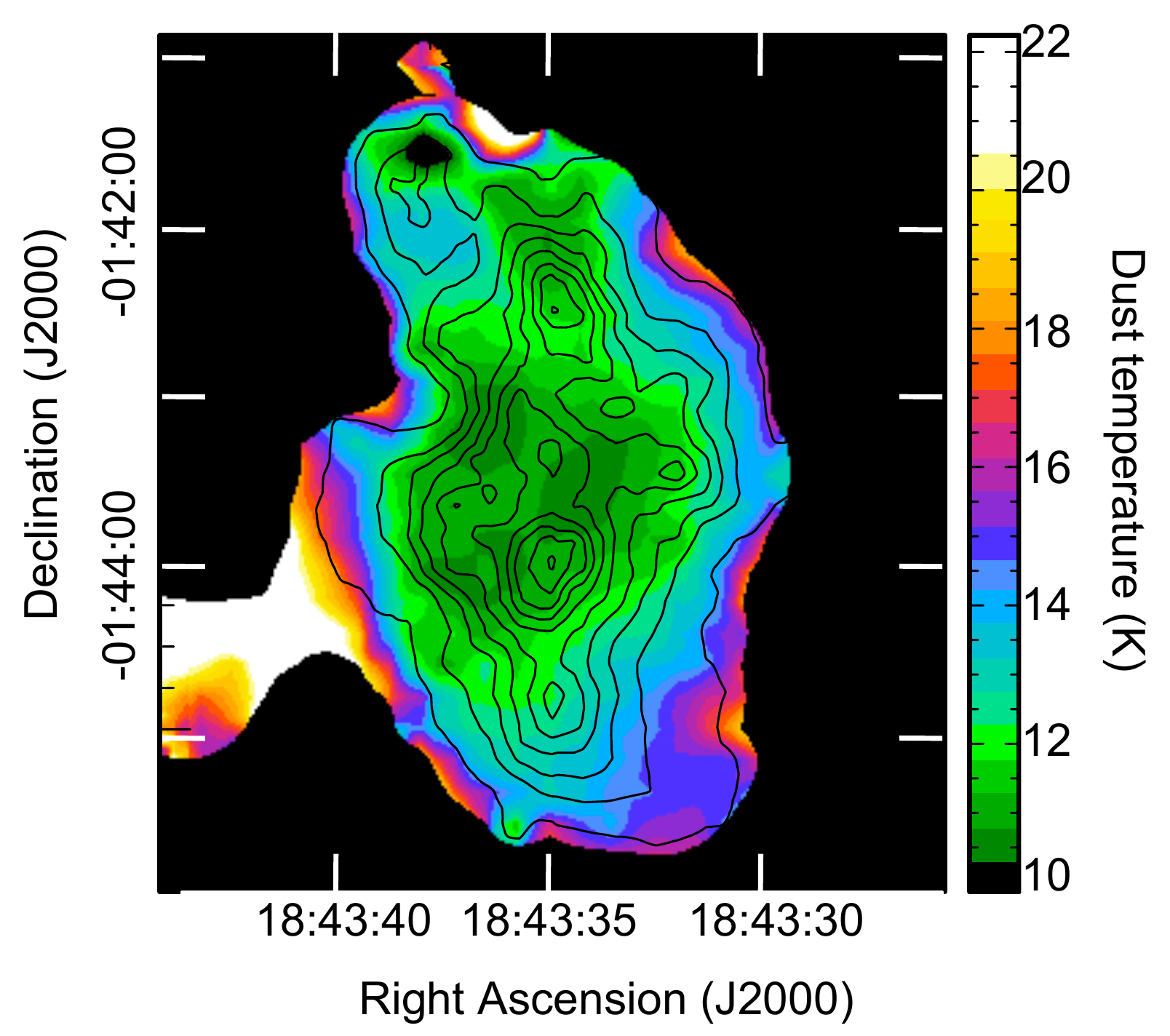}
%ss.eps}
%\vspace{-2cm}
\caption{Dust temperature map (color scale) for the IRDC shown in Fig.~\ref{sdc2col}. The contours represent the H$_2$ column density of this cloud, from $1\times10^{21}$ to $3.5\times10^{22}$~cm$^{-2}$. We see that the strongest peaks in the 250~\microns image of Fig.~\ref{sdc2col} are not the strongest column density peaks. 
\label{col}}
\end{figure}

The SED fitting for IRDC pixels was performed using 4 data points per pixel,
smoothed to the same resolution  of 36\arcsec. 
    The 70~\microns point  was excluded since at this wavelength the
  optically thin assumption is not valid, the IRDCs are usually seen in absorption and, therefore, the optically thin assumption is invalid
   (see Sect. 5 for a discussion about 70~\microns point like sources).  Because of the small number of data points for each fit, we fixed
  $\beta$  leaving two free parameters. Adopting the same value of $\beta=2$ for all IRDCs, we calculated dust
    temperature maps (cf Figs.~\ref{sed} and \ref{col} ).  The first
  important feature  of the map is that the dust
  temperature is non-uniform, ranging from 10~K for the coldest central cores,
  up to 22~K, consistent with the inferred background temperature. 

  To obtain a superior angular resolution, we iterated the SED
  fitting excluding this time the 500~\microns point, and using the
  dust temperature map and column density map calculated in the first
  iteration as an input parameter to the fit of the second iteration.
  Doing so, the temperature map changes little, while the column
  density map appears more structured. This is not really surprising
  because the temperature is predominantly constrained by the shortest
  wavelength data, toward the peak emission and on the exponential
  part of the Planck function. Given this, we directly used the
  36\arcsec~ resolution temperature map in combination with the
  250~\microns~images to obtain 18\arcsec~ resolution column density
  maps. The resulting column density map for the IRDC shown in
  Fig.~\ref{sdc2col} is shown as contours in Fig.~\ref{col}. We
  assumed a normalization of the specific dust opacity law of 0.12
  cm$^{2}$~g$^{-1}$ at 250~\microns, adopting the thin ice mantle dust
  grain models of \citet{ossenkopf1994} at a density
  n=10$^5$~cm$^{-3}$.  Uncertainties in this normalization and $\beta$
  directly implies at least a factor of 2 of uncertainty in the column
  density and a couple of degrees for the temperatures.  It is
    possible that $\beta$ decreases systematically towards the inner
    region of IRDCs because of dust coagulation \citep{ossenkopf1994}. However, even a decrease from $\beta=2$ to $\beta=1.5$ increases
    the minimum temperature by only $\sim2$~K, which is insufficient to reverse the observed trends discussed in the next section.
  %However, they do not change the general trends we discuss in the next section.
  
  %We will discuss the impact of choosing such a normalization in
  %Section~5.

  We can already note several important points here. First, the strongest
  peaks at 250~\microns seen in Fig.~\ref{sdc2col} are not the strongest
  column density peaks. Dust temperature variations 
    have an important role in determining the emission. Most of
     the column density peaks  are also correlated with local temperature minima, although there is not a one to one
     correlation. The 3D
  structure of the cloud is probably responsible for the apparent projection
  offsets. 
  %Finally, from the combination of the dust temperature map and
 % column density map, we can estimate the mass weighted temperature over the
  %entire cloud which is 12.1~K. 

%\begin{figure}[!t!]
%\hspace{-0.5cm}
%\vspace{-0.0cm}
%\includegraphics[width=9.5cm,angle=0]{sdc_temp.pdf}
%\caption{ Dust temperature map. We see temperature variations across the IRDC ranging from 22~K down to 10K, with a mass weighted temperature for the entire cloud of 12.1~K. It is worth noting that if we can identify clearly some local minima, their positions do not correspond exactly to the column density peaks observed in Fig.~\ref{col}.
%\label{temp}}
%\end{figure}

\section{Discussion}

\begin{figure}[!t!]
\hspace{-0.cm}
\vspace{-0.0cm}
\includegraphics[width=8.5cm,angle=0]{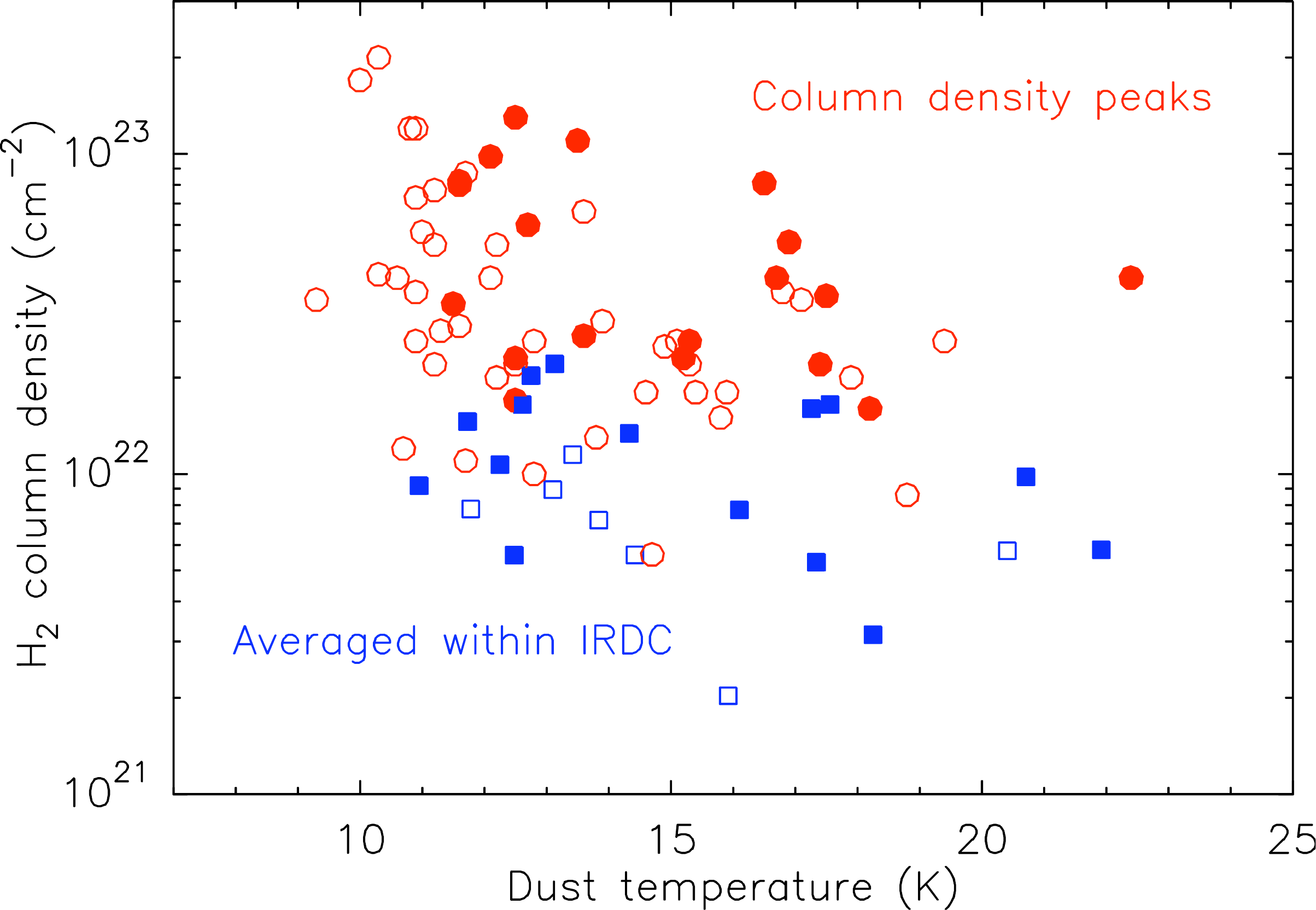}
\caption{Dust temperature versus column density plot. The blue square
  symbols show the average column density and mass weighted
  temperature for the 22 IRDCs analyzed. The red circular symbols show
  the temperature and column density at the column density peaks
  within the IRDCs. The filled symbols correspond to IRDCs with at least one 70~\microns point-like source, i.e., star-forming IRDC/peaks.
  \label{TvsN}}
\end{figure}

The most striking feature of our dust temperature maps is the non-isothermal nature of IRDCs. In Fig.~\ref{TvsN}, we plot the mass-weighted dust temperature versus the average H$_2$ column density where the blue symbols correspond to the 22 analyzed IRDCs, while the red symbols correspond to the dust temperature and H$_2$ column density peaks we visually identified within the IRDCs (for instance 6, peaks are identified in the IRDC of Fig.~\ref{col}).  While the median mass-weighted temperature is around 15~K, we see that it can 
be as low as 8~K in the inner part of IRDCs. Knowing that background temperature lies in a 20 to 30~K range, this means that the temperature decreases by 10 to 15~K from edge to center. Therefore, the dust temperature in dense molecular clouds is highly non-uniform.  

Temperature and its fluctuations are fundamental parameters for fragmentation in molecular clouds. Hydrodynamical simulations including radiative transfer \citep{bate2009,krumholz2010} showed that protostellar heating could significantly reduce fragmentation on small scales.  The picture we can draw from the study presented here suggests that IRDCs in the Galactic plane probably form from warm molecular gas with a temperature of $\sim20$~K, which cools efficiently to 10~K. On the other hand, in the previously mentioned simulations temperature gradients only develop once the first protostars form, in contrast to what we found here. 

The filled symbols in Fig.~\ref{TvsN} correspond to IRDCs (blue) or column density peaks (red) that contain at least one 70~\microns 
point-like source. We can see that two thirds of the IRDCs and one third of the column density peaks are currently forming stars, consistent with the statistics from Spitzer observations of a large sample of IRDCs  (Peretto \& Fuller, in prep). For these star-forming cores, the column density and
temperature are probably overestimated and underestimated, respectively,  since we did not take into account the data points at wavelengths below 100~\microns. However, since 
the longest wavelengths have more weight in the SED fitting (due to their lower uncertainties) the column density of the cold gas should not be affected too much.
Another interesting point is that when extrapolating the linear relation found by \citet{dunham2008} between the 70~\microns flux of low mass protostars and their bolometric luminosity to the 70~\microns sources we identified here, we find that their bolometric luminosity lies somewhere between a few 1000~L$_{\odot}$ and a few 10 000~L$_{\odot}$ depending on the source and their distance (typically between 3 and 6~kpc). This makes these sources   excellent high-mass protostellar candidates. 

Finally, Fig.~\ref{TvsN} clearly shows high column density fragments with no 70~\microns point-like sources within them.  These are interesting objects for high-resolution follow-up studies since they might well be the birth places of massive stars, before any star formation activity has started. Extrapolated to the entire Spitzer IRDC catalog of \citet{peretto2009}, we expect to find a few hundred similar starless high column density cores in the full Hi-GAL dataset.

\section{Conclusions}

We have developed a technique to obtain reliable temperature and column density maps of IRDCs using the first Herschel data from the Hi-GAL
open-time key project. We have demonstrated our method for a small sample of massive IRDCs by showing that they are non-isothermal but have a definite temperature structure, their inner regions being cooler than their edges. This temperature structure probably has an important impact on the fragmentation of the cloud in the early stages of its evolution. We also 
identified  a number of high column density peaks, reaching values of $10^{23}$~cm$^{-2}$, which are most likely to be the birth sites of massive stars. The full Hi-GAL dataset is expected to contain several hundreds of these sources.

\bibliographystyle{aa}
\bibliography{references}

\end{document}